\documentclass{IEEEtran}
\usepackage{graphicx}
\usepackage{subfigure}
\usepackage{amsmath}
\usepackage{balance}
\usepackage{float}
\usepackage[misc,geometry]{ifsym}
\usepackage{color}
\usepackage{url}
\usepackage{array}
\newcolumntype{P}[1]{>{\centering\arraybackslash}p{#1}}

\begin{document}
\title{Context-based Co-presence Detection Techniques: \newline A Survey}


\author{\IEEEauthorblockN{Mauro Conti\IEEEauthorrefmark{2}, and
Chhagan Lal\IEEEauthorrefmark{3}}\\
\IEEEauthorblockA{Department of Mathematics, University of Padua, Padua, Italy \IEEEauthorrefmark{2}\IEEEauthorrefmark{3}\\
Email:\IEEEauthorrefmark{2}conti@math.unipd.it,
\IEEEauthorrefmark{3}chhagan@math.unipd.it,}}

\maketitle
\begin{abstract}
\boldmath

In this paper, we present a systematic survey on the contextual information based proximity detection techniques. These techniques are heavily used for improving security and usability in Zero-Interaction based Co-presence Detection and Authentication (ZICDA) systems. In particular, the survey includes a discussion on the possible adversary and communication models along with the existing security attacks on ZICDA systems, and it reviews the state-of-the-art proximity detection techniques that make use of contextual information. These proximity detection techniques are commonly referred to as \textit{Contextual Co-presence} (COCO) protocols, which dynamically collect and use contextual information to improve the security of ZICDA systems during the proximity verification process. Finally, we summarize the significant challenges and suggest possible innovative and efficient future solutions for securely detecting co-presence between devices in the presence of adversaries. The proximity verification techniques presented in the literature usually involve trade-offs between metrics such as efficiency, security, deployment cost, and usability. At present, there is no ideal solution which adequately addresses the trade-off between these metrics. Therefore, we trust that this review gives an insight into the strengths and shortcomings of the known research methodologies and pave the way for the design of future practical, secure, and efficient solutions.
\end{abstract}

\begin{IEEEkeywords}
Relay attack, Zero-interaction authentication, Context-aware, Sensor modalities, Distance bounding, RFID, Proximity detection.
\end{IEEEkeywords}

\section{Introduction}
\label{sec:intro}
Nowadays, there are many industrial applications which grant specific services and privileges based on the physical proximity of the communicating devices. For instance, we use contactless smartkey to unlock our car, even to start the engine without inserting the key. These industrial applications use the most popular short-range communication technologies known as Radio Frequency Identification (RFID)~\cite{rfid} and Near Field Communication (NFC)~\cite{nfc}, for establishing contact between the communicating pairs. Other widely used applications that use short-range contactless smartcards based on RFID or NFC include supply chain management, e-passport~\cite{passport}, access cards (such as building, parking, highway toll fee collection and public transport~\cite{transport}), medical implants, Point-of-Sale (PoS) systems~\cite{mastercard}, to name a few. Moreover, the smartcard-based access control systems that require proximity verification and authentication are also being deployed in safety and security-critical infrastructures such as military research facilities and nuclear power plants. Therefore, it is essential to secure such systems against all types of adversaries. The main reason for the popularity of contactless authentication systems compared to contact-based smartcard systems is their higher overall user experience concerning ease in manageability and usability. However, due to the inherent weaknesses in underlying wireless communication, the RFID/NFC systems are exposed to a wide variety of security and privacy attacks~\cite{Akinyokun2017}. Thus, it subverts the security and usability advantages offered by these authentication systems. 

\par The so-called relay attacks are one of the many distance hijacking attacks that exploit the radio communication technology of RFID/NFC systems~\cite{Drimerr2007}~\cite{Aurelien2010}~\cite{Franciss2010}. In relay attacks, a proxy device (often referred to as \textit{ghost}) that emulates a contactless smartcard is placed near the reader to impersonate a victim's card within the proximity to the reader. On the other end of the communication point, a mole (often referred to as \textit{leech}) that acts as a reader is placed near to the victim card~\cite{Kfir2005}. Both of these malicious devices are in control of an adversary. The proxy forwards all the messages to the mole which act as a fake authentic reader for the victim card. The distance between the proxy and mole can be increased as far as the communication delay is kept sufficiently short. Example of instances that show the vast existence of relay attacks are demonstrated in~\cite{Sportiello2013}. Authors in~\cite{Sportiello2013} show a successful relay attack over more than 300 miles, and the authors in~\cite{Aurelien2010} demonstrate relay attacks on passive keyless entry and start for over 50 meters. Furthermore, in~\cite{Korakk2014}, the successful relay attacks over more than 110 meters are demonstrated using three NFC smartphones.

\par To overcome the above mentioned inherent vulnerabilities and attacks on contactless communication systems, various researchers are working towards different defense techniques. The proposed techniques try to preserve the fundamental properties such as zero-interaction and usability of the systems while ensuring the protection from the distance hijacking attacks. The two most commonly found defense techniques in the state-of-the-art are the contextual co-presence~\cite{[15]}~\cite{[33]} and the distance bounding~\cite{DBrands1994}~\cite{Bengio1991}~\cite{Beth1991} protocols. Both these protocols provide zero-interaction authentication~\cite{Corner2002} by using co-presence detection as an additional security measure on top of the basic authentication process.


\par In this paper, the co-presence detection techniques that are based on distance bounding protocols are considered out of the scope. It is because we aim to review only the applications that make use of resource-constrained (e.g., smartcards and smartkeys) and commodity devices (e.g., smartphones and tablets) as provers and verifiers. The distance bounding needs to be implemented at the lowest possible layer in the communication stack because even a small error in estimating processing time at the prover-side can lead to significant deviations in the distance bound. Therefore, implementing distance bounding on commodity devices like ordinary smartphones might be a challenge. However, we direct the interested readers toward the following comprehensive distance bounding research works~\cite{Brelurut2016}~\cite{avoineACMCS17}.      


\subsection{Motivation and Contributions}

Considering the higher potential damage such as an unauthorized entry in a secure and sensitive facility, stealing a car, credit card frauds, and skipping tolls, which could be caused by exploiting the vulnerabilities in co-presence systems. Thus, these systems require robust and secure authentication models. Over the years, researchers have proposed many solutions based on distance bounding and context-aware information protocols, which develop patches to fix the identified vulnerabilities. To the best of our knowledge, this is the first attempt which provides an extensive overview of the attacks and their prevention techniques for ZICDA access control systems. However, some efforts have been made to describe the problem and its possible solutions within one specific protocol such as distance bounding protocols for distance based attacks~\cite{Brelurut2016}, or within a particular communication technology like RFID~\cite{Jannati2015}. But, these state-of-the-art survey articles does not sufficiently cover the details of all the ongoing attacks and their proposed solutions on proximity-based systems. For instance, in~\cite{Hanck2009}, authors discuss the feasibility of implementation and the corresponding security implications for various active and passive relay attacks, and in~\cite{Poturalsk2011}, the authors present a brief survey on multiple attacks and their countermeasures using distance bounding protocols with IEEE 802.15.4a (i.e., Impulse Radio Ultra-Wideband). Additionally,~\cite{Hanck2009} and~\cite{Poturalsk2011} are outdated given the extensive research that has been done in the last few years on the security of co-presence systems. It is because the attack vector has been increased significantly in recent years due to the rapid deployment of zero-interaction systems in various real-world scenarios such as health-care, PoS, and keyless car entry. Hence, we firmly believe that a comprehensive survey is essential for an audience who plans to initiate their research work in this direction. Our paper does not attempt to solve any new challenges but presents an overview and discussion on security threats and their countermeasures in ZICDA systems. We believe that we have taken here the required initial steps that will help understand how to make full use of the contextual information to provide flexibility, and to strength decision making in access control systems. 

\par In this paper, we provide the first comprehensive survey on co-presence detection techniques. To this end, the major contributions of our work are as follows.

\begin{itemize}

\item We discuss the key security problems that affect the use of contactless smartcards in ZICDA systems. We review security threats, vulnerabilities, and attacks specific to these systems. In particular, we survey the literature over the period 2000-2018 by focusing our attention on the impact analysis of security attacks performed on ZICDA systems.

\item We present a general architecture for ZICDA system, which includes its characteristics, deployment challenges, and applications. Furthermore, we discuss the communication and adversary system models that are being used in ZICDA systems. In particular, we assist interested readers in understanding the existing challenges in the deployment of ZICDA systems, estimate the possible damages caused by the adversaries, and improve the techniques for proximity detection and containment processes. Furthermore, we provide an overview concerning feasibility, robustness, and effectiveness for the existing and potential attacks over ZICDA systems, and we examine the risks for users of these systems.

\item Finally, we present a survey of the state-of-the-art security solutions for detecting co-presence using contextual information (i.e., contextual co-presence protocols). We further extend our survey by including the co-presence detection techniques that also emphasize on the importance of privacy preservation (mainly regarding user location) during the access control authentication processes in Location-Based Services (LBS). Please note that in this paper only context-aware security solutions with respect to ZICDA systems have been considered for survey, and we have not surveyed the context-aware solutions that are used for improving the security and privacy of users in other application domains such as mobile applications~\cite{Zhauniaro2014}~\cite{Sikder2017}, Internet of Things (IoT)~\cite{Sezerr2018}, Industrial IoT~\cite{Bisio20188}, and future wireless networks~\cite{Nguyen20188}. Additionally, we discuss how the existing approaches ensure fundamental security requirements and protect communications in the ZICDA systems together with the open challenges and strategies for future research work in the area.
\end{itemize}


\subsection{Organization}
The rest of the paper is organized as follows. In Section~\ref{sec:Overview_C_CopII}, we present the overview of ZICDA systems, which include its characteristics and applications. In the same section, we also discuss details about communication model, authentication system, and adversary model used for proximity verification of communicating devices in a ZICDA system. Furthermore, at the end of the Section~\ref{sec:Overview_C_CopII}, we discuss all the existing attacks and their impacts on the ZICDA systems. In Section~\ref{sec:Cop_detIII}, we review existing solutions, which are proposed for detecting the co-presence between the communicating devices. We broadly discuss the proximity detection techniques that are based on contextual co-presence protocols. In Section~\ref{sec:Iss_FutureIV}, we present open issues and directions for future work. Finally, Section~\ref{sec:conclusionsV} concludes our work.

\section{Zero-Interaction based Co-presence Detection and Authentication}
\label{sec:Overview_C_CopII}
In this section, we present the overview of a generic functional model of Zero-Interaction based co-presence Detection and Authentication (ZICDA) access control system. First, we introduce the deployment techniques that have been used in a ZICDA system. Then, we discuss the standard communication and adversary model for ZICDA systems. 

\subsection{Overview of ZICDA system}

A ZICDA system represents a specific set of contactless access control systems in which the access-seeking entity (e.g., smartcard, smartkey, and smartphone) will implicitly prove their co-presence with the verifier along with its authentication credentials. For instance, Passive Keyless Entry (PKE) system, which is also named as ``Smartkey'' system is an automobile's electronic lock to its doors and ignition system. In PKE system, the driver carries a token (i.e., smartkey) that communicates (using RFID technology) with car's access control system to unlock the doors and activate the ignition, only if, the token's authenticity and proximity are successfully verified.
\par Verifying the proximity along with the authenticity is necessary for ZICDA systems. Otherwise, these systems become vulnerable to various type of Man-In-the-Middle (MIM) attacks such as eavesdropping, distance-hijacking, data corruption and manipulation, and relay attacks. One way to detect proximity is through received signal strength, but an adversary can easily manipulate signal strength through active relays. Authors in~\cite{Jannati2015} provide study depicting that different types of proximity-based access control systems are susceptible to MIM attacks. Mainly, the relay attacks are successful in ten car models from eight different vendors~\cite{Aurelien2010}. In addition to vehicular systems, these attacks can easily target credit/debit cards and smartphones, which uses NFC technology and contactless smartcards.

\subsection{Communication Technologies}
In ZICDA systems, due to the resource-constrained nature of prover (e.g., smartcard and smartkey), the following three short-range and low-energy sensor technologies are commonly used for communication between prover and verifier: (i) Radio Frequency Identification (RFID), (ii) Near Field Communication (NFC), and (iii) Bluetooth Low Energy (BLE). Among these three, NFC is used in a large array of applications because it combines the security of BLE with the short-range data transfer capabilities of RFID. For NFC to work, one must tap or wave the smartphone (acts as NFC reader) against an NFC tag to secure an object's context or to perform an action. For example, you could purchase chocolates just by tapping your NFC reader against the box of chocolates. NFC-capable devices such as a smartphone can work as a reader as well as a tag. It is predicted that NFC will be used as a key technology in realizing the Internet of Things (IoT)~\cite{Parada2017} paradigm. It is due to the enhanced security features of NFC such as a user can easily pair an NFC tag with another form of authentication on hand (like the license in your wallet) to create a two-pronged authentication system. The above feature is particularly relevant in the health-care world~\cite{Wazid2017}, and it is even being mandated by the Drug Enforcement Administration (DEA) as a standard security practice.
 
\par NFC is generally viewed as a finely honed subset of Radio Frequency Identification (RFID). NFC operates at the same frequency (i.e., 13.56 MHz) as high-frequency RFIDs, and it performs many of the similar operations as RFID tags (and readers) and contactless smartcards. The NFC can operate in the following communication modes. 
\begin{itemize}
\item Read/Write: In this mode, an NFC-enabled reader/writer device (such as a smartphone) can read information from the smart objects, and act upon the received information to improve services provided by these smart objects. By performing a simple touch of these devices to the smart objects, the users can perform various tasks such as short message service (SMS) texts without typing, automatically connect to websites via a retrieved URL, and get information about various relevant offers or obtain coupons. This mode is beneficial for realizing the Internet of Things (IoT) services.     
\item Peer-to-Peer: In this mode, one NFC-enabled reader/writer device can communicate with another NFC-enabled reader/writer device. One of the reader/writer devices behaves as a tag to create the communication link.

\item Card emulation: While working in this mode, an NFC-enabled reader/writer device can replace a contactless smartcard, which enables NFC devices to be used within the existing smartcard infrastructure for services such as making payments at PoS, access control at building entrance or for a vehicle, toll gates, and medical implants.
\end{itemize}

\par As NFC is a subset of RFID, the standards and protocols for NFC are based on RFID standards as outlined in FeliCa, ISO/IEC 14443~\cite{Issovits2011} and some are parts of ISO/IEC 18092. These standards govern the use of proximity cards using RFID technology.

\begin{figure*}
\centering
  \includegraphics[scale = 0.35]{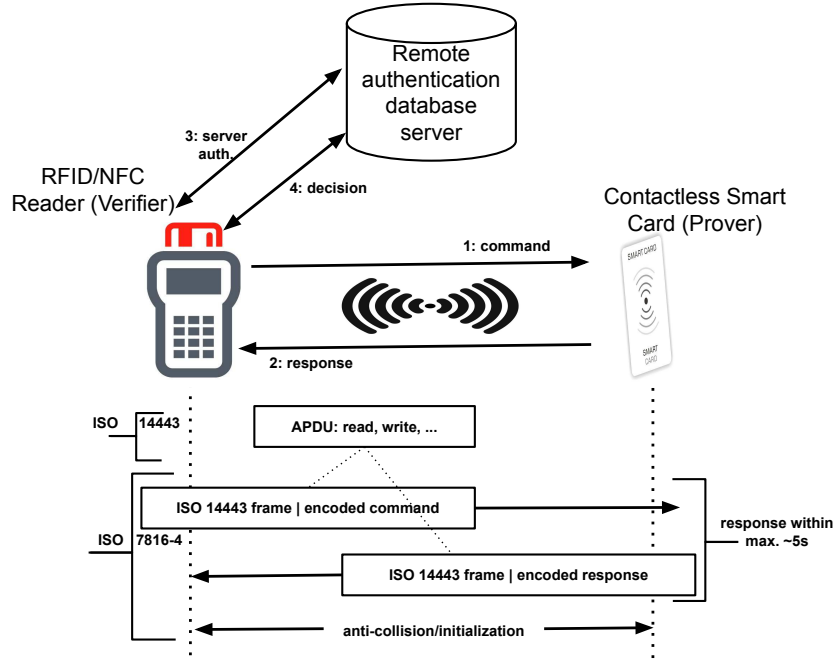}
\caption{\small{Communication model for ZICDA systems}}
  \label{Fig:2}
\end{figure*}

\subsection{Communication and Adversary Models}

Figure~\ref{Fig:2} shows a generic communication model for ZICDA systems. The communication model consists of two devices namely prover ($P$) and verifier ($V$). To get access to the system, $P$ has to authenticate itself to $V$ and also prove that $P$ is in close proximity to $V$. The authentication process between the devices, i.e., $P$ and $V$, triggers automatically when both devices are nearby to each other. The communication traffic between $P$ and $V$ is encrypted using a pre-shared secret key, which is generated using either shared-key or private/public key model. The $P$ encrypt its authentication information using the secret key before transmitting it to $V$. Depending upon the application and system implementation, a ``credential verification'' function make the authentication decision for $P$ at $V$ either locally or remotely as shown in Figure~\ref{Fig:2}. For example, in a PoS application, a user (i.e., $P$) performs the contactless payment using her NFC-enabled smartphone at a PoS terminal. In this specific application, the ``credential verification'' function is stored at the web server of a bank whose credit card is being used for the payment at PoS terminal. Other applications such as locking/unlocking a car using a smartkey, where the ``credential verification'' function is integrated with the terminal device itself. In ZICDA access control systems, the smartcard (i.e., a user token such as an access card, key or mobile phone) acts as $P$, and the terminal (i.e., a desktop computer, wall-mounted device or car system) plays the role of $V$. 

\begin{figure*}
\centering
  \includegraphics[scale = 0.35]{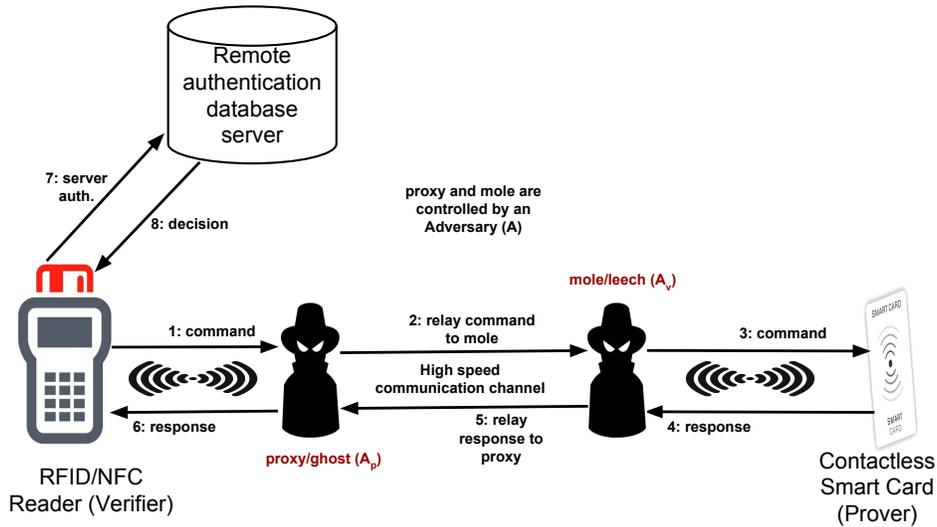}
\caption{\small{Adversary model for ZICDA systems}}
  \label{Fig:3}
\end{figure*}

\par The adversary model for ZICDA systems is shown in Figure~\ref{Fig:3}. We assume that an adversary possesses following standard Dolev-Yao~\cite{Dolev1981} features and capabilities: a) adversary ($A$) has complete control over the communication channel used for authentication process between $P$ and $V$, and b) $A$ has no physical access or possession of $P$ and $V$, nor can $A$ compromise the functionality of $P$ and $V$ devices. Therefore, none of the benign entities in the communication protocol of ZICDA system can be tampered with or compromised. However, $A$ is allowed to stay close to $V$ and $P$. The main aim of $A$ is to fool $V$ into concluding that $P$ is in proximity. 

\par Figure~\ref{Fig:3} illustrates that an adversary ($A$) which resides in the proximity of $P$ and $V$ can perform various types of distance hijacking (i.e., distance-reduction or distance-enlargement) attacks on the wireless channel between $P$ and $V$. First, $A$ can intelligently place one device, called mole ($A_v$), in the proximity to $P$ without $P$ knowing about it. Then, $A$ places another device called, proxy ($A_p$), close to $V$ which emulates a contactless smartcard. Both $A_v$ and $A_p$ communicate using a high bandwidth channel. In this way, $A$ takes the form of a ``mole-and-proxy'' (or often called ``ghost-and-leech'') duo ($A_v$, $A_p$), and it relays messages to and forth between $V$ and $P$. This process leads $V$ to conclude that $P$ is in proximity and vice-versa. Therefore, such a simple adversary model can fully compromise the security and privacy of an ordinary ZICDA system without requiring any physical access to the communicating devices nor does it requires the authentication credentials.           

\subsection {Attacks vector for ZICDA Systems}
\label{sec:attacks}

Due to technological advancements in mobile devices and radio frequency communications, a broad array of applications such as contactless payments, keyless entry systems, smart posters, to name a few, are deployed rapidly for mass-market users. These applications use contactless authentication along with the proximity verification between the communicating devices for ensuring secure access control. The increased overall user experience regarding ease in manageability and usability are the main attractions of these applications. Unfortunately, the radio channel used for communication is vulnerable to various security and privacy attacks such as eavesdropping~\cite{Hancke2011}, relay attack~\cite{Hancke05}~\cite{DCavdar2015}, impersonation~\cite{Avoine2009}, and distance hijacking~\cite{Cremers2012}. Thus, these attacks limit the usability of co-presence techniques in various application domains. To provide security against all types of attacks in ZICDA systems is a challenging task. Ideally, a ZICDA system should be protected against the following attacks.

\begin{itemize} 
\item \textit{Mafia fraud:} \textit{Mafia fraud} attack~\cite{Boureanuu2015}, also called relay or wormhole attack is first introduced by~\cite{Conway21976} and~\cite{Chun2006}. In this attack, the $V$ and $P$ are honest and far apart, and an adversary tries to shorten the physical distance between them. The adversary uses a similar attack scenario as described in Figure~\ref{Fig:3}, the attacker places a proxy verifier ($V'$) near $P$ and a proxy prover ($P'$) near $V$. These proxies create an extended high bandwidth communication link between $V$ and $P$ by relaying all the communication messages between them. In this way, $P'$ and $V'$ make $V$ and $P$ to falsely conclude that both are in close proximity. Traditional cryptographic-based security techniques cannot prevent the \textit{Mafia fraud} attacks because the proxies (i.e., $P'$ and $V'$) need not to perform any decryption or encryption on communication messages nor they require to run any authentication process with $V$ and $P$. Thus, these proxies can create an effective, transparent communication link between $V$ and $P$. This attack has been successfully demonstrated in various ZICDA Systems in which NFC/RFID techniques are used for communication between $V$ and $P$.       
\begin{figure}
\centering
  \includegraphics[scale = 0.42]{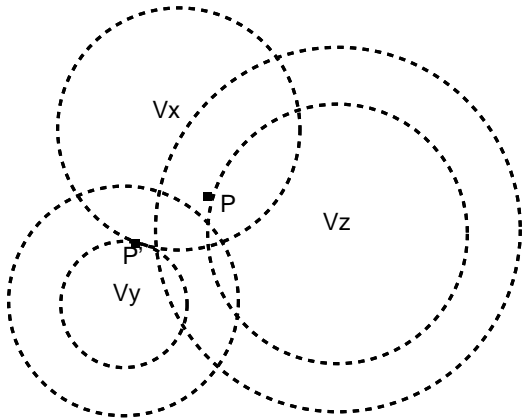}
\caption{\small{Distance fraud attack}}
  \label{Fig:4}
\end{figure}

\item \textit{Distance fraud:} In \textit{Distance fraud} attacks~\cite{DBrands1994}, a sole fraudulent prover ($P'$) convince the honest verifier (V) that she is at a different (usually shorter) distance than she really is. Unlike \textit{Mafia fraud} attack, here the prover itself is dishonest, and only the verifier is a victim. \textit{Distance fraud} attacks are most effective and disastrous for real-time location-based systems (RTLS)\footnote{RTLS are automated systems that determine the locations of assets.}. Application instances of RTLS include manufacturing, logistics, and supply chain management, where expensive components or parts of a final product and other important entities involved are being tracked throughout the whole logistics process. A practical example of how \textit{Distance fraud} attacks can adversely affect the RTLS is shown in Figure~\ref{Fig:4}. In this scenario, three nodes (i.e., $Vx$, $Vy$ and $Vz$) perform continuous tracking of the current location of the node $P$ using its received signal strength. We can see from Figure~\ref{Fig:4} that if node $P$ wants to be malicious, it could pretend to be at position $P'$ at the same time when it is at $P$. To perform this action, $P$ decreases its signal strength when communicating with node $Vz$ while it increases the signal strength when communicating with $Vy$. In this case, the verifier nodes $Vy$ and $Vz$ are unable to detect the fraud of $P$ because she is a legitimate node, and she authenticates herself with true credentials.           

\item \textit{Terrorist fraud:} A slightly different version of \textit{Distance fraud} attack in which a dishonest prover ($P'$) attacks the system with the help of a third party attacker ($A$) is called \textit{Terrorist fraud}~\cite{Vaudenay2013}. In \textit{Terrorist fraud} attack, the $P'$, which is far apart from an honest verifier ($V$), conspires with $A$, who is close to $V$ to masquerade as the honest prover by providing $A$ with selected credentials for authentication. Let's consider an example, assume that $A$ is a terrorist who wants to cross the border. $P'$ helps $A$ in answering the questions of the immigration officer (i.e., $V$). Another example could be the one in which $A$ help $P'$ in applications such as location forging. Assume a scenario involving electronic monitoring using an ankle bracelet. \textit{Terrorist fraud} attack enables the subject (i.e., $P'$) of the electronic monitoring system (i.e., $V$) to leave her residence with the help of $A$ who stays close to $V$. 

\begin{figure}
\centering
  \includegraphics[scale = 0.33]{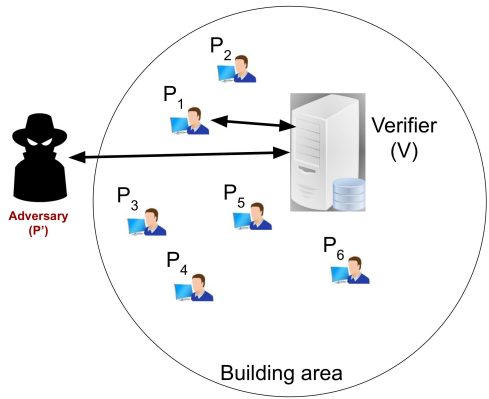}
\caption{\small{Distance-hijacking attack}}
  \label{Fig:5}
\end{figure}

\item \textit{Distance-hijacking:} In Distance-hijacking attack~\cite{Cremers2012}, a dishonest far-away prover ($P'$) exploits one or more honest close-by provers' $\{P_{1}, P_{2}, \cdots, P_{n}\}$ to provide a verifier $V$ with false information about the distance between $P'$ and $V$. Consider a real-world scenario as shown in Figure~\ref{Fig:5}, in which several employees (i.e., $\{P_{1}, P_{2}, \cdots, P_{n}\}$) work in a secure building. A mainframe system (i.e., $V$) containing sensitive information is located inside the building. Any authorized employee can get access to $V$ through their contactless smartcard. To complete the authorization process with $V$, an employee needs to be in the building along with her valid credentials. Now, assume that an adversary ($P'$), which has a (stolen) smartcard is sitting outside the building along with a powerful antenna. To access $V$, $P'$ already has the valid security credentials, but $P'$ also need to prove that she is inside the building. For this purpose, $P'$ performs eavesdropping over the communication channel of the distance bounding protocol, which is running between the employee $P_1$ and $V$. Distance bounding works in two phases; in the first phase, the $P$ needs to prove to $V$ that both are in proximity to each other. After successful completion of the first phase, $P$ authenticates itself to $V$ using valid credentials. To perform distance-hijacking attack, $P'$ jam the communication link between $P_1$ and $V$ as soon as the first phase of the distance bounding is completed. Then $P'$ will complete the second phase on behalf of $P_1$ using her (stolen) credentials. In this way, $V$ now believes that the $P'$ is in the building with valid credentials, thus she is granted the access.             

\item \textit{Location cheating~\cite{WHe2011}:} It is a colluding attack in which a close-by helper and a far-away dishonest prover ($P'$) collude to prove that $P'$ is close to verifier ($V$). Location-based services (LBS) led by foursquare\footnote{http://www.foursquare.com}, GasBuddy\footnote{https://www.gasbuddy.com/}, GyPSii\footnote{http://www.gypsii.com}, Loopt\footnote{http://www.loopt.com}, and Dark Sky\footnote{https://darksky.net/app/} has attracted a lot of attention in recent years. The LBS uses the geographical position of a user to enrich user experience in a variety of contexts such as location-based searching and location-based mobile advertising. To attract more users, the location-based mobile social networking services provide rewards and offers to the user when it checks into certain venues or locations. This gives incentives to users to engage in location-cheating for their benefits. Dishonest provers may obtain undeserving benefits at specific venues (i.e., places like coffee shops, restaurants, shopping malls, to name a few) by making multiple false location check-ins at different times. \\ For example, Foursquare connect users to local businesses like shops or restaurants by using their current location information. Many business owners offer concrete benefits such as free vouchers, special offers, and cash rewards to the most active registrants visiting their shops or restaurants. In such a scenario, a $P'$ can perform location-cheating attack by taking help from her friend sitting in or near a restaurant. The close-by helper of $P'$ will use the credentials of $P'$ and prove her presence along with the authentication to trick the $V$. A vast array of LBS services use GPS locations that can be obtained from a user's smartphone. In such services, a user performs location-cheating~\cite{WHe2011} by exploring the open source operating systems of smartphones (e.g., Android) to modify global-positioning-system-(GPS)-related application programming interfaces (APIs). Once tempering is done, a user can cheat on her location using falsified GPS information.  

\end{itemize}


\section{Context-aware Co-presence Detection Techniques}
\label{sec:Cop_detIII}

In this section, we present a comprehensive survey of existing context-based co-presence detection techniques that address one or more security threats discussed in Section~\ref{sec:attacks}. The basis of provisioning contextual security in ZICDA systems is the fact that all devices residing in the proximity with each other will always ``see'' (nearly) the same physical and ambient environment (i.e., availability of suitable context). With the recent advancements in the hardware of mobile devices, these devices are now equipped with one or more inbuilt ``sensors'' such as microphones (for audio), wireless networking interfaces (for WiFi connectivity), global positioning system (for location), Bluetooth (for short-range communications), and other physical environment sensors (humidity, gas, temperature and pressure/altitude). The data collected using these sensors can be used as supplemental information to improve security decisions in ZICDA systems. With the information extracted from these sensors, the security decisions can be taken dynamically at the time the decisions are made. For this purpose, during the authentication process in ZICDA systems, two honest communicating devices can exchange and compare the dynamically gathered supplemental information to determine their co-presence towards each other.

\subsection{System Model for Context-based Access Control in ZICDA Systems}
\begin{figure}
\centering
  \includegraphics[scale = 0.33]{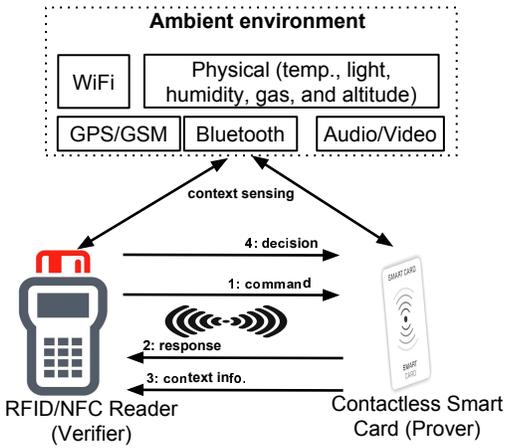}
\caption{\small{Context-based co-presence Detection System}}
  \label{Fig:7}
\end{figure}

Figure~\ref{Fig:7} depicts the generic system model for ZICDA systems that are based on contextual co-presence detection techniques. The main aim of most of the context-based ZICDA systems is to provide security against relay attacks. We can see from Figure~\ref{Fig:7} that the $P$ and $V$ will ``see'' (almost) the same ambient environment if they lie in close proximity. When either one of them leaves their common ambient environment, the context gathered by $P$ and $V$ will not match during the co-presence detection process, and the access to the system will be denied (please refer to Figure~\ref{Fig:7}).  

\par The general working methodology framework for ZICDA system models is shown in Figure~\ref{Fig:7}. The framework functions in two phases, in the first phase, when a $P$ enters in the transmission range (aka proximity) of a $V$, $P$ sends a trigger to $V$. Once triggered, $V$ start the authentication process with $P$ by sending one (or more) challenge(s) ($ch$) to $P$, upon reception of $ch$, $P$ generate (using her private key) a response ($rsp$) and send the $rsp$ back to $V$. After successful authentication, in the second phase, $V$ and $P$ initiate a context sensing process for a pre-defined set of contexts for a fixed duration of $t$. The context information collected by $P$ within duration $t$ can be represented by a vector ($\rho$) such that $\rho = \{ \rho_1, \rho_2, \rho_3 \cdots \rho_n$\}, where $n$ is the number of sensor modalities used to form the contextual information. Similarly, $\nu = \{ \nu_1, \nu_2, \nu_3 \cdots \nu_n \}$ is the corresponding vector of $n$ sensor modalities collected by $V$. Based on the similarity index calculated using vectors $\rho$ and $\nu$ at $V$, the access of $P$ to the system is either allowed or denied. The effectiveness and correctness of the calculations for the similarity index at $V$ depends upon the feature extraction, classification, and machine learning methods used in the process. The use of contextual security in the authentication process not only improves the security of the system but also provides flexibility in access control decisions. 

\par A point worth mentioning here is the ``initial delay'' in authentication process incurred due to the use of contextual security. Due to this, a trade-off arises between system access delay and its usability, i.e., a high delay will lead to lower usability and vice-versa. This initial delay can be minimized to some extent using the following approaches: (i) reduce the number of context during the context-aware authentication. However, it will decrease the level of security provided by the system, and (ii) $V$ can perform the credential and contextual authentication processes for $P$ in parallel, but if $P$ uses a low battery power device for authentication then this method posses a high energy consumption and $P$ needs to perform frequent recharges, thus it reduces the usability of the system.

\begin{figure*}
\centering
  \includegraphics[scale = 0.32]{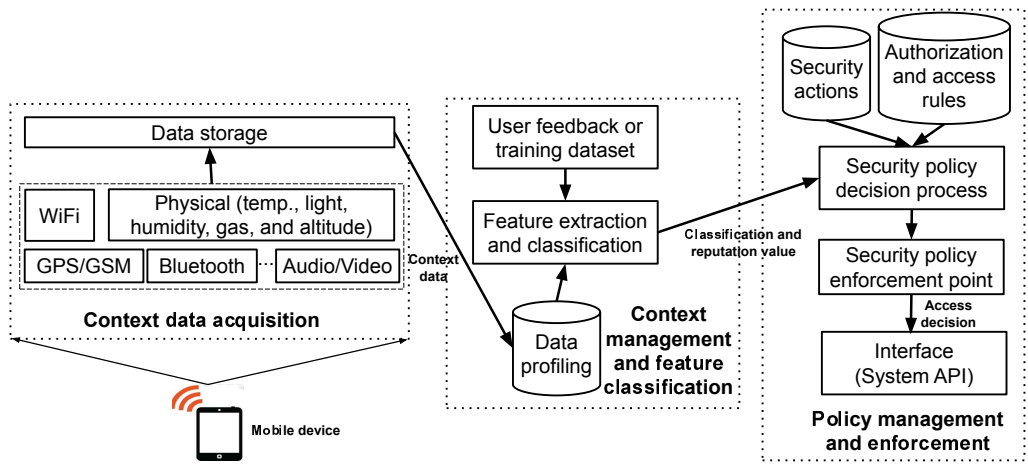}
\caption{\small{Overview of a Generic Contextual Co-presence Security Framework}}
  \label{CC_Framework}
\end{figure*}

\subsection{Context-based Co-presence Detection Framework}
\label{sec:Con_aware}


The basic requirements to achieve contextual security in any access control system are (i) availability of contextual information, (ii) efficient integration of available contextual information in runtime, and (iii) instant availability of contextual information to security analysts. In practice, a typical major obstacle in incorporating context into a security monitoring program is the availability of the contextual information in a format that supports integration with log and alert data. Additionally, the contextual information needs to be validated to ensure that it is accurate and has integrity. The ideal platform integrates the data and information in real or near real-time to allow not just the linkage of the data and knowledge efficiently and effectively, but it also enables rules and complex event processing that occurs due to the use of contextual information. Finally, the platform must have the capability to make the integrated information quickly and readily available to security analysts to present a ``scenario'' (as discussed in the earlier contactless smartcard examples) that provides all of the information required to validate, respond to, and mitigate incidents.

\par Figure~\ref{CC_Framework} shows the interaction between the major components involved in a generic Contextual Co-presence Security Framework (CCSF). The existing proximity detection techniques either use the whole or parts of the CCSF to verify the co-presence between the communicating devices. The CCSF apply context profiling and machine learning algorithms on real-world reference dataset that is collected in an uncontrolled environment, and it evaluates the effectiveness of automatic and adaptive context classification for detecting co-presence. For an access control system, the CCSF architecture is used for training the classifier using the ground truth data. The trained classifier will then used as a context comparator to compare the contextual data that is received from a prover and verifier at runtime. The CCSF can be instantiated depending upon the requirements and applications of the underlying access control system. We can see in Figure~\ref{CC_Framework} that the CCSF mainly consists of three major components namely, context data acquisition, context management, and feature classification, and policy management and enforcement. Below, we briefly discuss the functioning and interactions among these three components.

\begin{itemize}
\item \textbf{Context data acquisition:} The CCSF architecture is driven by the contextual information collected by this component. The accuracy of the assessment of co-presence detection depends highly on the data collection and aggregation process used by the data acquisition module. The contextual security refers to the use of additional information (i.e., context) to improve the security at the time when security decisions are made. Therefore, the context data sensing is done dynamically at the time when a decision must be made for an access control system. Depending upon the type of the application, the communicating entities involved in the context-aware authentication process needs to gather a set of predefined contextual data types using their corresponding inbuilt sensors. The sensed data by the verifier and the prover devices is then compared to verify the co-presence between them. The comparator must be trained, in advance, using the ground truth data (or reference data) to provide precise interpretation, analysis, and decision making. The task of collecting the ground truth data is the main aim of the context data acquisition component of CCSF. Once collected, the reference data is passed to the next component of CCSF (called context management and feature classification) for training the classifier.

\par To gather contextual data, one can install an easy-to-use and secure application on a large number of user devices. The user devices involved in the data collection process could be smartphones, tablets, or a specific purpose device such as~\cite{Sensordron}. Data collection is a critical phase of the framework because the number and quality of the ground truth data that has been collected have a high impact on the accuracy of the assessment of co-presence detection. Data collection is a time-consuming, expensive, and cumbersome process. To collect the reference data, an onsite assert deployment is required. The data collection could be done using the dedicated users (employees or suppliers) or an approach such as crowd-sourcing, where data collection is done by soliciting contributions from a large group of people (self-selected volunteers or part-time workers). In both cases, the users will have to carry the data collection device with required sensing capabilities (hardware or software). The former approach is expensive, and the collected dataset will be of small size, but the data will be trustworthy and of high quality. While, the latter will collect data that is inexpensive and it will have higher data quantity, but it will be less secure and of low quality. In particular, to build a robust and accurate CCSF, the collected reference data should have characteristics such as high quality and quantity, accuracy, timelinesses, and variability. Furthermore, there are strict government issued guidelines (to support user or information privacy) such as General Data Protection Regulation (GDPR) that needs to be followed during the process of data acquisition. Collecting real-world reference data to train the classifiers and update them periodically is one of the biggest challenges in context-aware access control systems. Once the training phase is complete and CCSF is ready to use, it is deployed in the corresponding real-world applications. After CCSF system deployment, the data collection component will collect sensor data dynamically, only at the time when a prover asks access for the access control system. 

\item \textbf{Context management and feature classification:} During the context-aware authentication phase the predefined contextual data is collected by prover and verifier devices, and it is sent to the context management and feature classification (CMFC) module. The CMFC module consists of three components: data profiler, classifier, and training dataset. Data profiling will help in quickly and thoroughly unveiling the true content and structure of the observed context data. The profiler will perform completeness, uniqueness, values distribution, range, and pattern analysis on received data to ensure that it is of adequate quality. Once analyzed properly, the profiler identifies the most promising features (i.e., feature selection) to build a feature vector describing the current context of the users.

\par The classifiers such as decisions trees, Support Vector Machine (SVM), and K-Nearest Neighbours can be trained under supervised learning using a reference dataset. Once the training phase is over, the classifier uses the context feature vectors, which are generated by the profiler to classify new observations (i.e., sensor data) concerning the current applications security and privacy-related properties. The classifier outputs the classification estimates and its associated confidence value, and these are forwarded to policy management and enforcement component, which considers them while making access control decisions. The performance of the classifier will directly influence both, security and usability, of the underlying access control system. In particular, the security of a system is determined by the False-Positive rate (i.e., erroneously established co-presence between far-away devices), while the usability is represented by the False-Negative rate (i.e., incorrectly established non-co-presence between nearby devices).        

\item \textbf{Policy management and enforcement:} This component provide final decisions on the on-going context-aware authentication process between prover and verifier devices. The policy management module uses the in-built policies along with the received confidence value to enforce suitable policies on the verifier. Depending on the type and number of policies enforced by this module, different users can receive varying levels of security access on the same access control system. For instance, a server can be only accessed if the user is within its proximity, but once access to the server is granted, different users can have different rights on the functionalities of the server. In such a scenario, the proximity check is coupled with the individual users' security policies, and therefore, the policy enforcement at the time of context-aware authorization is required.                  
\end {itemize} 

\par In the next section, we will discuss the state-of-the-art context-aware co-presence detection techniques that utilize, partially or wholly, multiple components from our above-discussed CCSF architecture.   

\begin{table*}
\caption {Context-aware Co-presence Detection Techniques - I}
\centering
\scalebox{1.4}{
\begin{tabular}{|P{1.0cm}|P{1.8cm}|P{2.2cm}|P{1.2cm}|P{1.3cm}|P{2.7cm}|}\hline
\textbf{Proposals} & \textbf{Communication channel} & \textbf{Sensor modalities} & \textbf{Privacy preservation support} & \textbf{Application specific} & \textbf{Description}\\ \hline

\cite{[1]} & RFID & magnetometers, accelerometer, GPS  & No  & No  & design context-aware selective unlocking mechanisms and secure transaction verification\\\hline

\cite{[2], [5]} & RFID/NFC & audio, ambient light  & No  & No  & determine the proximity by correlating certain sensor data extracted from the two devices\\\hline

\cite{[3]} & RFID & GPS  & Yes  & PoS  & location-aware secure transaction verification scheme\\\hline

\cite{[4]}~\cite{[28]} & RFID & audio  & No  & PKES  & sound-based proximity-detection method\\\hline

\cite{[6]} & RFID/NFC & WiFi (radio waves)  & No  & No  & authenticate co-located devices based on their shared radio environment\\\hline

\cite{[7]}~\cite{[32]} & RFID/NFC & temperature, single-bit round-trip  & N/A  & No  & elliptic curve-based mutual authentication protocol \\\hline

\cite{[8]} & Bluetooth /RFID/NFC & audio, WiFi, and GPS  & No  & No  & comparing and fusing different sensor modalities in ZIA systems\\\hline

\cite{[9]}~\cite{[12]} & Bluetooth /NFC & ambient noise and luminosity & No & IoT domains & secure ZIA pairing suitable for IoT and wearable devices\\\hline

\cite{[10]} & N/A & GPS, WiFi  & Yes  & mobile applications & detection against device misuse and sensory malware\\\hline

\cite{[11]} & N/A  & audio and luminosity  & N/A  & proofs-of-presence (PoPs) & solutions against context guessing attacks in LBS\\\hline

\cite{[13]}~\cite{[31]} & RFID/NFC & accelerometer, gyroscope  & N/A  & access control systems  & authorized reference trajectories on Transparent Authentication (TA) schemes
\\\hline

\cite{[14]} & WiFi  & trajectory through a road network (gyroscope signal, and GPS) & Yes  & VANETs & technique to verify the ongoing co-presence of vehicles in an urban environment
\\\hline
\cite{[15]} & RFID/NFC & WiFi, Bluetooth, GPS, and audio)  & No  & No  & investigate the performance of different sensor modalities for co-presence detection
\\\hline

\end{tabular}}
\label{T:1}
\end{table*}

\begin{table*}
\caption {Context-aware Co-presence Detection Techniques - II}
\centering
\scalebox{1.4}{
\begin{tabular}{|P{1.0cm}|P{1.8cm}|P{2.2cm}|P{1.2cm}|P{1.3cm}|P{2.7cm}|}\hline

\textbf{Proposals} & \textbf{Communication channel} & \textbf{Sensor modalities} & \textbf{Privacy preservation support} & \textbf{Application specific} & \textbf{Description}\\ \hline

\cite{[16]}~\cite{[18]} ~\cite{[19]}~\cite{[20]} & Bluetooth/ RFID/NFC  & artificial ambient environments (infrared light, sound, etc) & N/A & time-restricted contactless transactions  & evaluated the effectiveness of 17 ambient sensors \\\hline

\cite{[17]} & N/A & bidirectional sensing and comparing button presses and releases behaviour & N/A  & EMV contactless payments  & detection based on sensing button presses on the user's smartphone by both transaction devices
\\\hline

\cite{[21]} & Bluetooth  & magnetometer & N/A & device pairing  & pairing smartphones by exploiting correlated magnetometer readings\\\hline

\cite{[22]}~\cite{[25]} \cite{Mayrhofer2007} & NFC & accelerometer & No & Yes & PoS and context-based technique to prevent mafia attack in mobile NFC payment \\\hline

\cite{[23]} & NFC  & audio and light & Yes & payment cards  & secure proximity detection techniques\\\hline

\cite{[24]} & N/A  & electromagnetic signals & Yes & LBS & privacy-preserving proximity testing\\\hline

\cite{[26]} & RFID/NFC & ambient temperature, precision gas, humidity, and altitude & No & PoS  & use of purely ambient physical sensing capabilities in authentication systems
\\\hline

\cite{[27]} & RFID/NFC  & Features-fusion~\cite{[8]}, and decisions-fusion & No & No  & systematic assessment of co-presence detection in the presence of context-manipulating attacker\\\hline

\cite{[29]} & RFID/NFC  & speech recognition, and location sensing/ classification & Yes & payment systems & defend against unauthorized reading and relay attacks\\\hline

\cite{[30]} & RFID  & features-fusion and decisions-fusion based on majority voting & Yes & ETC systems  & unauthorized reading and relay attacks detection in RFID ETC systems \\\hline

\cite{[20]} & WiFi, Bluetooth, infra-red & accelerometer & No & smartcard & continuous  two-factor authentication \\\hline

\cite{[33]}~\cite{[34]} & Bluetooth  & audio & No & online banking  &  a usable and deployable two-factor authentication mechanism \\\hline

\end{tabular}}
\label{T:2}
\end{table*}

\subsection{Co-presence Detection using Contextual Information}


In this section, we discuss the co-presence detection techniques that use contextual information during the proximity verification process to improve the security of the target access control system. The contextual information is collected using one or more sensor modalities that reside in the prover device(s). In recent years, the use of mobile devices for spontaneous communications increases significantly in various applications. Therefore, securing these communications from multiple attacks such as relay attack, eavesdropping, and impersonation becomes a vital precondition. For instance, an attacker can read, relay, and modify messages between communicating peers without either peer suspecting that the communication between them has been tempered. The use of contextual security as an additional layer on top of the traditional security can help to prevent such malicious third party attacks, as apparently, no user want their private information being leaked or tampered with. The primary motivation for the researchers to develop context-aware solutions is the rapidly ongoing technological and hardware advancements that enable many RFID/NFC tags to be equipped with many low-cost sensing capabilities. Over recent years, sensors with various sensing capabilities have been incorporated in RFID tags~\cite{Yeager2009}~\cite{Sample20099}. For instance, Intel's Wireless Identification and Sensing Platform (WISP)~\cite{Sampless2007}~\cite{Smith22006} has developed tag with various sensing capabilities, and this extends the use of RFID beyond simple identification. With the help of these advanced RFID devices, one can efficiently provide numerous promising applications for pervasive sensing and computation. It also paves the way towards providing improved security and privacy services by leveraging contextual information from the existing physical environment. Tables~\ref{T:1} and~\ref{T:2} depicts the state-of-the-art context-aware co-presence detection techniques along with their short description and some other related information such as the sensor modalities used for content gathering, the communication channel(s) considered between the prover and verifier, and the support provided by a proposed approach is specific to an application, or it is generic, and the support for user privacy (where applicable) is provided or not.        

\par The use of contextual information to improve the security of access control applications is not a new technique. For example, the banking authentication system uses time and location as contextual information to provide an additional security layer in online transactions. In this scenario, assume a customer that wants to transfer all her funds to a third-party account. The transaction appears genuine, i.e., the customer has authenticated itself correctly to the bank, she is accessing an account for which she is authorized, and the third-party bank account appears valid too. However, the access location or time of the transaction looks suspicious, e.g., the account has been accessed from a location which is far from the home location of the customer or the activation time of the transaction is not consistent with the previous transactions timestamp pattern of the customer. Therefore, without the additional context, the bank is unable to determine if the activity is fraudulent or not. 

\par The use of contextual information to improve the security of access control systems has rapidly increased in recent years, and it is mainly due to the advancements in the mobile device and communication techniques, which makes the availability of the content more accessible to these systems. In~\cite{WuC2008}, the authors propose an approach that provides additional security using context in role-based access control (RBAC) systems. In particular, the main aim is to combine contextual security (by using location and time as context) and role-based access control to retail business processes, which uses the RFID technology for inter-communication. Furthermore, in~\cite{Covington2002}, authors propose a context-aware security architecture for emerging applications, and in~\cite{Seo2010} a context-aware remote security control for mobile communication devices has been proposed. In both these works, the contextual information such as location, time, and network access points (like WiFi) is used to improve security. It is done by dynamically setting the security policies for individuals based on their current threat levels.

\par In~\cite{[3]} and~\cite{[30]}, authors propose context-based security techniques that uses onboard tag sensors to collect contextual information (location and speed). The proposed techniques minimize the likelihood of unauthorized reading and relay attacks in RFID Electronic Toll Collection (ETC) and banking access control systems. In~\cite{[30]}, the context data sensed through GPS sensors is used to develop a context-aware selective unlocking technique for tags at ETC such that they can selectively respond to reader challenges.  

\par In~\cite{[6]}, authors propose a proximity-based authentication technique called ``Amigo''. To authenticate co-present mobile devices, Amigo uses knowledge of their shared radio environment as proof of physical proximity. The key advantages of Amigo include the following: (i) it does not require any additional hardware, (iii) it does not require user involvement in the authentication process, and (iii) it is not vulnerable to eavesdropping. The main idea is that the co-present devices will simultaneously monitor a common set of ambient radio sources (WiFi access points or cell phone base stations) to perceive a similar radio environment. An evaluation conducted using WiFi-enabled laptops show that Amigo is robust against a range of passive and active attacks. To further strengthen the fact that co-present devices will see the common radio environment, fluctuations in the signal strength of existing ambient radio sources are considered in~\cite{Varshavsky2007}. It shows a reduction in false positives and false negatives in the system. 

\par In~\cite{Krumm2004}, authors present a system called NearMe. NearMe discovers what is already nearby and to augment context for ubiquitous computing. For this purpose, NearMe server determines proximity by comparing a list of WiFi access points and signal strengths called ``WiFi signatures'' from its clients. To use NearMe, each client has to perform the following three functions: (i) register with proximity server, (ii) report recent WiFi signature, and (iii) query nearby places and peoples. A similar proximity testing system which uses WiFi access points and Bluetooth signals to generate ``location tags'' is introduced in~\cite{[24]}. The system was implemented and evaluated on the Android platform. Along with security, it also guarantees the privacy preservation for the clients involved in it.     

\par Based on the audio and light data collected from the ambient sensors that are available in NFC enabled smartphones, a secure proximity technique is presented in~\cite{[23]}. The main aim is to prevent relay attacks at point of sale (PoS) systems, where just bringing the NFC enabled smartphone close to PoS is sufficient to complete a transaction. In particular, authors propose a transaction verification mechanism that can determine the proximity (or lack thereof) between honest verifier and prover by comparing specific sensor data (audio or light), which is extracted from the communicating devices. In~\cite{[28]}, a secure radio channel between communicating devices based on similar audio patterns has been proposed to develop an unobtrusive but cryptographically strong security mechanism. Furthermore, in~\cite{[29]}, authors use ambient audio for secure device pairing on android mobile phones. In this work, audio is used as a metric to generate a secure cryptographic key that establishes communication between mobile distributed devices.

\par The use of Secret Handshakes as context information rapidly increases in a large array of applications that uses RFID or contactless cards for access control purposes. The intuition behind its use as the content is as follows. Let's assume a typical usage scenario such as RFID or contactless card-based entry in a secure facility. When a prover wishes to enter in an access-controlled building, she often subconsciously (thus, increases the usability of the system) does a fixed set of motions such as her left/right-hand reaches for her wallet, draws her purse out or wave it near the door's reader, and take a pause. From the above use-case, one can observe if it is possible for the RFID chip or contactless technology in the access cards to somehow internally detect a pattern, which depicts precisely when, how, and in what order these actions were being performed. If this is the case, then it is possible to install appropriate logic on the RFID tags and contactless cards that would only allow access when these actions are matched.

\par Based on the secret handshakes mechanism, several techniques have been proposed over the years to combat various distance-based attacks (please refer to Section~\ref{sec:attacks}). To detect and prevent Man-In-The-Middle (MITM) attacks, authors in~\cite{Mayrhofer2007} propose a technique for device-to-device (e.g., phone and handset) authentication that includes an additional layer in a traditional security suite. It is done by adding additional information in terms of shared movement patterns. A user can simply generate various patterns by shaking the devices together, and these patterns can be easily captured using accelerometer sensors that are embedded in the communicating devices. In particular, two methods that combine cryptographic primitives with accelerometer data analysis are proposed to establish secure radio channels by creating authenticated secret keys. Further, in~\cite{[2]}, authors propose context-aware mechanisms to defend against the RFID unauthorized reading (by using owner's posture recognition as context information) and relay attack reading (by using audio as context information). Similarly, in~\cite{[31]} and~\cite{[8]}, authors propose gesture and motion recognition based techniques to defend against ghost-and-leech (a.k.a. proxying, relay, or man-in-the-middle) attacks in RFID tags and other contactless cards. All these techniques increase the resilience of access control systems against a set of proximity-based attacks. However, the issue of user privacy caused by the gesture recognition process which involves the sensitive user data (i.e., user's biometric information) is not adequately addressed in these works. 

\par For the first time, authors in~\cite{[15]} systematically investigate the impact of using a single as well as a set of sensor modalities on proximity detection systems. First, a standard data collection and processing framework similar to the one we have described in Figure~\ref{CC_Framework} is developed. The proposed framework runs in realistic everyday setting to collect data, which is then used to train the classifier. Second, the authors compare the performance of four commonly available sensor modalities (i.e., WiFi, Bluetooth, GPS, and audio) using various combinations, first individually and then in the sets of two and four. The work provides a comparison regarding resisting relay attacks in zero-interaction based access control systems for each combination. The authors claim that WiFi data as the context is better in opposing relay attacks when compared to other sensor modalities, and the fusing of multiple modalities further improve resilience against relay attacks. However, the fusion of various modalities retains a high level of system usability up to a certain point. We argue that with the increase in the number of sensor modalities, the time required to authenticate a prover and the complexity of context-aware algorithms deployment increases. Thus, it decreases the usability and feasibility of an access control system. In~\cite{[15]}, to make the proximity detection techniques more robust and versatile, the authors motivate the need for a stronger adversarial model in which the adversary can compromise the integrity of context sensing mechanisms. For instance, an attacker can create fake Wifi access points, add random noise, and it can modify the purely ambient physical sensing capabilities~\cite{[26]} such as ambient temperature, precision gas, humidity, pressure, and altitude.

\par One of the most comprehensive works towards analyzing, extending, and systematizing state-of-the-art tasks on context-aware proximity detection under a stronger, but a realistic adversarial model is presented in~\cite{[27]}. In this work, authors present a systematic assessment of proximity detection in the face of context-manipulating adversaries. It has been shown that not only the content manipulation is possible, but an attacker can consistently control and stabilize the values of multiple, heterogeneous (e.g., acoustic and ambient physical environment) sensors using low-cost, off-the-shelf equipment. Thus, an attacker who can manipulate the context gains a significant advantage in defeating access control systems that are based on contextual security techniques.

\par Authors in~\cite{[33]} propose a representative approach called \textit{Sound-Proof}, a usable two factor authentication that leverages ambient sound to detect co-presence between the \textit{phone} (used as a second authentication factor) and the \textit{browser} (a login terminal such as a banking website) running on a different mobile device such as laptop or tablet. Sound-Proof claims to found an optimal trade-off between the usability (i.e., it does not require an interaction between the user and her phone) and security (i.e., secure login on a browser in the presence of remote attackers). In particular, Sound-Proof uses the audio signatures collected from the microphones of the two devices. Sound-Proof provides a useful security enhancement on top of the traditional password-only authentication technique that is commonly used to perform online banking transactions. The only essential requirement in Sound-Proof is that the user should keep her phone near to the laptop while doing the login tasks. However, a weakness of the Sound-Proof is identified by authors in~\cite{[34]}. In~\cite{[34]}, authors show that to perform an attack, the remote attacker does not have to predict the ambient sounds near the phone as assumed in the Sound-Proof, instead, it can deliberately make or wait for the phone to produce predictable or previously known sounds (e.g., ringer, notification or alarm sounds). Therefore, exploiting the weakness as mentioned above, a full attack system can be launched to compromise the security of Sound-Proof successfully. 

\par Authors in~\cite{Han2017} aims to authenticate messages in VANETs through physical context comparison. The physical context consists of the surface of the road that includes road conditions such as bumps and potholes which can be measured using the accelerometer. Later, the context is used to derive a secret key, which is shared between the co-present vehicles. However, the entropy of the context to generate the secret key and the effect of different road surfaces remains unexplored. Thus, it makes the security guarantees of the system unclear. Recently, authors in~\cite{[14]} propose an approach to verify the ongoing co-presence between two vehicles in an urban environment. The method exploits the characteristics of a trajectory (using gyroscope signals, GPS, etc.) through a road network. The aim is to allow authenticity checks for safety-critical applications. The approach requires a vehicle to share the same route as a leading vehicle to become a verified following vehicle. Co-present vehicles gain knowledge of verified neighbors as well as the capability to authenticate their VANET messages. The construction only reveals a driver's trajectory to other co-present vehicles, and therefore, it protects passengers privacy against an eavesdropping attacker. The proposed approach operates transparent to pseudonym schemes, and thus, it cannot be exploited to attribute different messages to the same sender. The proposal has been implemented as an Android application to evaluate its performance in experiments involving two cars. 

\subsection{Co-presence Detection in Location-Based Services}

One of the primary goals of pervasive computing is to build service applications that are sensitive to the user's current context information. For example, location-based apps such as Swarm, Foursquare, Glympse, and Google-now, which uses the user location as a context to dynamically provide various services (e.g., information of nearby places, friends, and shops). One way to provide such services is to determine proximity by measuring absolute locations and compute distances. However, computing perfect location threatens user privacy, and it is also not necessarily easy to calculate, especially indoors, where GPS on user devices does not work well, which is usually a place where people spend most of their time. These Location-Based Services (LBS) use the approximate geographical position to enrich user's Quality of Experience (QoE) concerning various contexts such as location-based searching and location-based mobile advertising. To attract more users, service providers give real-world rewards to the user when it does check-in at a specific venue or location. These rewards motivate users to cheat on their real locations. In particular, LBSs can be defined as an array of services available with mobile devices (e.g., smartphones, tablets, and smart-watch), tailoring their functionality to current positions or trajectories of users or vehicles~\cite{Wang20187}. 

\par In~\cite{WHe2011}, authors investigate vulnerabilities leading to possible location cheating attacks in LBS applications and discuss possible countermeasures for the same. By using Foursquare as a use-case scenario, a new location cheating attack is proposed, which can easily cheat the current location verification techniques. The paper shows that if an attacker carefully studies the open-source operating systems for mobile devices such as Android to modify GPS-related application programming interfaces (APIs), then the attacker can cheat their location by altering the GPS information. While LBSs offer great opportunities for a large array of customer-oriented services, but at the same time, it also presents significant privacy threats to the users. To strengthen the mechanisms for preventing location-cheating in LBS, authors in~\cite{[11]} propose Proofs of Presence (PoP) based resilient techniques against malicious users. The paper present facts indicating that the use of context-aware PoPs for verification of users' location claims is vulnerable to \textit{context guessing attacks}. Furthermore, it proposes two countermeasures to mitigate \textit{context-guessing attacks}. The first countermeasure called ``surprisal filtering'' is based on profiling and estimating the entropy associated with individual PoPs. The second countermeasure suggests the use of longitudinal observations of ambient physical properties of the context. In~\cite{Shin20122}, the authors investigate and discuss the trade-off issues between users' location privacy protection and their Quality of Service (QoS) for the LBSs. 

\par The basis of LBS comes from spatial and temporal big data, which is provided by an enormous amount of mobile devices through GPS and various communication networks (e.g., cellular networks and WiFi). Using LBS to perform co-presence detection poses a significant threat to user privacy. To address this issue, various privacy preservation LBS schemes have been proposed in recent literature. For example, the authors in~\cite{Wang20187} first investigate the privacy issues in LBSs concerning possibilities of sensitive data leakage and then propose an approach that preserves query data intending to provide accurate LBS answers with zero-server-knowledge on query data. In most of the state-of-the-art schemes for privacy preservation in LBS, a single trusted anonymizer is placed between the users and the location service provider (LSP). However, it limits privacy guarantees and incurs high communication overhead when used in continuous LBSs. It is because once the anonymizer is compromised, it may put the user data at risk. Authors in~\cite{Zhangs2017} propose a dual privacy preserving technique for continuous LBSs to protect the users' trajectory and query content privacy. In this approach,  multiple anonymizers are placed between users and LSP, which are combined with Shamir threshold mechanism, dynamic pseudonym mechanism, and K-anonymity technique. Similarly, to achieve an adequate balance among user privacy, usability, and efficiency in LBSs, authors in~\cite{CDI2017} proposes SPOIL, which is a practical location privacy approach for LBSs. In particular, the idea is that a client (i.e., mobile device) shifts user-intended point-of-interests (POIs) to some neighboring POIs and query the mapping server using the shifted POIs. 

\section{Open Issues and Directions for Future Work}
\label{sec:Iss_FutureIV}

In this section, we present the lessons learned from our survey that includes an array of security threats to the ZIA systems, and the state-of-the-art context-based co-presence detection techniques that have been proposed to improve the security and privacy of various applications which uses these systems. Additionally, we discuss open issues and directions for future work that could lead to possible improvements in securing the ZICDA systems.


\par Based on our survey, the context-aware co-presence detection is emerging as a promising approach for defense against the relay attacks, which is considered as a significant threat to ZICDA systems. In context-based co-presence detection techniques, the contextual information is gathered from the surrounding environments that mainly includes audio-radio environment (e.g., ambient audio, WiFi, Bluetooth, infrared, and GPS, and combinations thereof) and physical environment (temperature, pressure, humidity, gas and altitude, and combinations thereof). Apart from contextual information based techniques, the distance bounding (DB) protocols have also shown significant potential in resisting various distance-hijacking attacks~\cite{Dimitrakakis2015}~\cite{Yang20188}. However, the use of distance bounding protocols in resource-constrained devices such as sensors, low-end smartphones, and smartcards are not suitable~\cite{Boure2015}. It is due to the multitude of hardware components and the multi-process architecture that is being used to implement the distance bounding techniques, which leads to unpredictable performance behaviour. In particular, these protocols interact at the physical layer, thus, the dedicated hardware is mandatory for practical implementations. Therefore, widespread deployment of DB protocols must await manufacturer endorsement.     

\par In the state-of-the-art, various types of context information that could be extracted from different sensor modalities such as magnetometers, accelerometer, GPS, and gyroscope, is used as contextual information to discover the co-presence between prover and verifier. Researchers have used single or a set of sensor modalities to generate some contextual signature that when matched up to a given threshold, the prover and verifier are considered within each others proximity. It can be deduced from the surveyed techniques that the use of multiple modalities provides more resistance to relay attacks and higher accuracy (i.e., lower false positives and false negatives). However, as the number of modality increases in the contextual set, the usability of the system decreases and the cost of the deployment increases. Additionally, the availability of the multiple sensor modalities depends on the device capabilities and the surrounding environment where the co-presence is being checked.      

\par Despite the availability of a broad array of context-aware co-presence detection techniques, the various types of distance-hijacking attacks still threaten the secure and efficient deployment of different emerging applications, e.g., secure message exchange in VANETs~\cite{Han2017}, two-factor authentication for user identification~\cite{BBasu}, and secure device pairing and service creation in IoT~\cite{Miettinen20144}~\cite{deMatos2017}. The correct implementation and functionality of these applications are based on the concept of contextual co-presence. Below, we discuss the challenges and future research directions that require significant research attention to improve the security of ZICDA systems and the privacy of its users.

\begin{itemize}
    \item \textbf{Integration of proximity proofs:} To integrate context-based co-presence schemes in a target system, the first requirement is the availability of the adequate context. However, having the context availability alone is not sufficient, it should be in the correct format or mechanism, and it should be validated to ensure that it is accurate and has integrity. Once the contextual information is available in an accurate, up-to-date, and validated format, it needs to be integrated based on key values, and a platform is required that enables the log. Additionally, alert data to be linked together with the contextual information is required to allow for efficient integration of contextual information in real-time. Finally, the system must have the capability to make the integrated information quickly and readily available to security analysts to present a ``scenario'' that provides all of the information required to validate, respond to, and mitigate possible security-related incidents. Performing the integration is easy if the system components (e.g., prover and verifier) only need software updates, but in cases where hardware updates are required, the development of the appropriate infrastructure is necessary. In particular, how to deploy contextual co-presence detection solutions cost-effectively and efficiently remains a research problem to address for future researchers.   
    
    \item \textbf{Usable solutions:} The use of contextual information is rapidly increasing in various application domains which include financial (e.g., Point-of-Sale, and multi-factor authentication for online and offline transactions) as well as non-financial (e.g., supply chain management, smartcard-based access, and medical implants) applications. Therefore, high usability becomes an essential requirement. However, the use of multiple sensor modalities for context gathering not only increases the cost of deployment, but it also decreases the usability of the system, which directly effects the quality-of-service (QoS) perceived by the end-users. Therefore, selecting an optimal yet minimal set of sensor modalities that effectively consider the tradeoff between the security, cost, and usability of the system remains an open issue.
    
    \item \textbf{Resistance against context manipulations:} Most of the state-of-the-art co-presence detection or relay attack resistance mechanisms consider the simplest adversary model (i.e., Dolev-Yao systems), hence these mechanisms might not be able to defend the system in the presence of an active adversary (i.e., context manipulating attackers). The existing research shows that it is trivial to modify, consistently control, and stabilize the context data gathered from different (single or multiple) audio-radio and physical environments using low-cost, off-the-shelf equipment~\cite{[27]}. Therefore, extensive research is required to ensure the robustness of the access control systems against distance hijacking attacks. For instance, the classifier and machine learning algorithms that are being used to train the system should consider the possibility of a strong adversary during the training phase. Also, the size of the training data set should be large enough, and it should exhibit the characteristics of the real-world data. 
   
    \item \textbf{Privacy preserving proximity detection:} In most of the available co-presence detection approaches, the context information consists of sensitive user data such as location, audio, and behavioural patterns. Therefore, it is essential to ensure the use of such contextual information in a privacy-preserving manner. However, it is hard to ensure privacy in co-presence systems due to the need for precise information that these systems require to perform the co-presence evaluation. For instance, it is hard to use the partial GPS information~\cite{Heng2016} and still do an accurate evaluation for co-presence. Hence, novel solutions are required to ensure privacy preservation during proximity detection.    
\end{itemize}

\section{Conclusions}
\label{sec:conclusionsV}

When a user tries to access a system, one can simplify the security decisions by basing it on binary choices (i.e., Yes or No). However, for the rapidly increasing thefts against logging credentials that are caused by the human or the system related errors, such binary decisions are not enough to protect the system. Therefore, if the verifier can base the security decisions on the who, when, where, when, what, and why behind the user's access request, it can develop usable security and privacy solutions for users without sacrificing the level of protection. This paper examines several ways that make use of a context-aware model (a new and adaptive security model), which feeds additional information to the Security Analytic Engine (SAE) to create efficient and flexible security decisions. In this paper, we start with the discussion on various real-world applications (e.g., PKE systems, contactless smartcard-based access control systems, contactless payment systems, inventory management, medical implants, and e-passport) and security threats (relay attack, terrorist fraud, location cheating, and impersonation) with respect to the ZICDA access control systems. We provided a comprehensive survey that includes all the state-of-the-art context-based co-presence detection techniques along with their merits and limitations. With the set of future research directions and challenges that we have discussed, we hope that our work will motivate fledgling researchers towards tackling the security, usability, and privacy issues of ZICDA systems.

\section*{Acknowledgements}
This work is supported in part by EU LOCARD Project under Grant H2020-SU-SEC-2018-832735. The work of M. Conti was supported by the Marie Curie Fellowship through European Commission under Agreement PCIG11-GA-2012-321980.

\bibliographystyle{IEEEtran}
\bibliography{copresence_det}

\end{document}